\documentstyle[11pt]{article}
\def\hybrid{\topmargin 0pt      \oddsidemargin 0pt
	\headheight 0pt \headsep 0pt
	\textheight 9in         
	\textwidth 6.25in       
	\marginparwidth .875in
	\parskip 5pt plus 1pt   \jot = 1.5ex}

\catcode`\@=11
\def\marginnote#1{}
\newcount\hour
\newcount\minute
\newtoks\amorpm
\hour=\time\divide\hour by60
\minute=\time{\multiply\hour by60 \global\advance\minute by-\hour}
\edef\standardtime{{\ifnum\hour<12 \global\amorpm={am}%
	\else\global\amorpm={pm}\advance\hour by-12 \fi
	\ifnum\hour=0 \hour=12 \fi
	\number\hour:\ifnum\minute<10 0\fi\number\minute\the\amorpm}}
\edef\militarytime{\number\hour:\ifnum\minute<10 0\fi\number\minute}

\def\draftlabel#1{{\@bsphack\if@filesw {\let\thepage\relax
   \xdef\@gtempa{\write\@auxout{\string
      \newlabel{#1}{{\@currentlabel}{\thepage}}}}}\@gtempa
   \if@nobreak \ifvmode\nobreak\fi\fi\fi\@esphack}
	\gdef\@eqnlabel{#1}}
\def\@eqnlabel{}
\def\@vacuum{}
\def\draftmarginnote#1{\marginpar{\raggedright\scriptsize\tt#1}}

\def\draft{\oddsidemargin -.5truein
	\def\@oddfoot{\sl preliminary draft \hfil
	\rm\thepage\hfil\sl\today\quad\militarytime}
	\let\@evenfoot\@oddfoot \overfullrule 3pt
	\let\label=\draftlabel
	\let\marginnote=\draftmarginnote
   \def\@eqnnum{(\theequation)\rlap{\kern\marginparsep\tt\@eqnlabel}%
\global\let\@eqnlabel\@vacuum}  }

\def\numberbysection{\@addtoreset{equation}{section}
	\def\theequation{\thesection.\arabic{equation}}}

\def\underline#1{\relax\ifmmode\@@underline#1\else
	$\@@underline{\hbox{#1}}$\relax\fi}

\def\titlepage{\@restonecolfalse\if@twocolumn\@restonecoltrue\onecolumn
     \else \newpage \fi \thispagestyle{empty}\c@page\z@
	\def\thefootnote{\fnsymbol{footnote}} }

\def\endtitlepage{\if@restonecol\twocolumn \else  \fi
	\def\thefootnote{\arabic{footnote}}
	\setcounter{footnote}{0}}  
\catcode`@=12
\relax

\def\beq{\begin{equation}}
\def\eeq{\end{equation}}
\def\bea{\begin{eqnarray}}
\def\eea{\end{eqnarray}}
\def\bar{\overline}
\def\z{\bar {z}}
\def\nn{\nonumber}
\def\pa{\partial}
\def\d{{\cal D}}
\def\p{{\cal P}}

\def\m{\mu}

\def\m{\mu}

\def\d{\delta}

\relax
\numberbysection
\hybrid
\begin{document}
\begin{titlepage}
\begin{center}
 \hfill    PAR--LPTHE 95--56  \\
 [.5in]
{\large\bf B-V QUANTIZATION IN 2-D GRAVITY AND\\ NEW CONFORMAL FIELDS}\\[.5in]
        {\bf   Laurent Baulieu }\footnote{email address:
baulieu@lpthe.jussieu.fr} \\
    	   {\it LPTHE\/}\\
       \it  Universit\'e Pierre et Marie Curie - PARIS VI\\
       \it Universit\'e Denis Diderot - Paris VII\\
Laboratoire associ\'e No. 280 au CNRS
 \footnote{ Boite 126, Tour 16, 1$^{\it er}$ \'etage,
        4 place Jussieu,
        F-75252 Paris CEDEX 05, FRANCE}
\\
and\\
RIMS, Kyoto University\\
 Kyoto, 606-01 Japan

\end{center}

\vskip .5in

\begin{quotation}
\noindent{\bf Abstract }      We investigate the properties of $2-D$ gravity in
the Batalin and
Vilkovisky quantization scheme. We find a   factorized structure which exhibits
duality properties analogous to those existing  in the topological theories of
forms.  New conformal field  are introduced with  their   invariant action.
\end{quotation}
 \end{titlepage}
 \newpage

%
\def\vap{\varphi}
\def\oz{{\overline{z}}}
\def\TB{\widetilde B}
\def\TG{\widetilde G}
\def\CD{{\cal D}}
\def\CF{{\cal F}}
\def\CB{{\cal B}}
 \def\vap{\varphi}
\def\oz{{\overline{z}}}
\def\TB{\widetilde B}
\def\TG{\widetilde G}
\def\CD{{\cal D}}
\def\CF{{\cal F}}
\def\CB{{\cal B}}
\def\be{\begin{eqnarray}}
\def\ee{\end{eqnarray}}
\def\pa{\partial}
\def\z{\bar{z}}
\def\p{{\pa_z}}

\def\m{\mu^z_\oz }
\def\mb{\mu^\oz_z}
\def\c{c^z}
\def\cb{c^\oz}

\def\nn{\nonumber}

\def\nn{\nonumber}
\newpage\null
 \section{Introduction}
In a previous paper, we have shown that the  Batalin and
Vilkovisky (B-V) formalism \cite{BV} for the gauge theories of forms coupled to
Yang-Mills   fields can be formulated in  a  unifying  algebraic framework
where
 the fields and the anti-fields are assembled into dual pairs
\cite{previous}. We have obtained a powerful    algorithm which generates
topological  actions of the Chern-Simon  and Donaldson-Witten type
  on the basis of vanishing curvature conditions.

In this paper we extend our analysis  to conformal theories coupled to 2-D
gravity.  We  show in particular that the   ghost system and
the  Wess and Zumino action of 2-D gravity  possess  a
structure    similar to that of a Chern-Simon   action and  we point out   the
possibility of introducing     new   conformal fields specific to 2-D
gravity, with an invariant action. This action  has a new gauge symmetry which
complements the ordinary conformal invariance.  Our work uses    the Beltrami
parametrization   which  gives a quasi Yang-Mills structure to the gauge
invariances of conformal theories  and   preserves the
factorization properties between the  holomorphic and anti-holomorphic
sectors.

The paper is organized as follows.  We first briefly  review  the
description of conformal  2-D gravity  with  Wess and Zumino terms in the
framework of the Beltrami  parametrization.   Then we   incorporate these
results in the Batalin and Vilkovisky quantization
scheme and find    a  duality picture between fields and anti-fields similar to
the one we had found in
\cite{previous}  for the theories of forms. All relevant conformal fields and
anti-fields, including the new fields   suggested by our unification procedure,
build up  dual pairs and  can be assembled as the components of
differential forms   with  a grading equal to  the sum of the ghost number and
of the  ordinary form degree.

\section{  2-D gravity in the Beltrami parametrization}

$2-D$ gravity has some very interesting algebraic properties when the metric is
expressed in terms of the Beltrami differential \cite{bellon}. Let us
briefly recall the  known  results. The Beltrami parametrization of the metrics
in 2-dimensional space   means that one expresses the length  of line elements
on
the worldsheet as
\begin{equation} ds ^2 = \Lambda (dz+\mu^z_{\oz} d\oz)(d\oz + \mu^\oz_zdz)
\end{equation}
 $\Lambda(z)$ is the conformal factor, $\m$ is the Beltrami differential   and
$z$ denotes the
complex coordinates.

This parametrization  has the following advantages. It permits one
 to build a
  theory which includes
the   Wess and Zumino field related to the conformal anomaly and
never refers to
the conformal factor
$\Lambda(z)$. Moreover, one can  define
a tensor calculus which only refers to the
Beltrami parametrization
\cite{bellon}
\cite{grimmnicolai}.

The BRST transformation laws of the Beltrami
differential $\mu^z_\oz(z)$ and of its ghost $\c(z)$
 take the following factorized form
\begin{eqnarray}
\label{brs} s\mu^{z}_\oz &=&\partial_\oz c^{z}+c^z\partial_z\mu^z_\oz-
\mu^z_\oz\partial_zc^z\nonumber\\ sc^z&=&c^z\partial_zc^z
\end{eqnarray} The relation between the Beltrami ghosts $\c(z)$ and $c^\oz(z)$
and the
$2D$-reparametri\-zation vector field  ghost $\xi^\alpha$ is \cite{bellon}
 \begin{eqnarray} c^z&=&(\exp i_\xi-1)(dz+\mu^z_\oz d\oz)=
 \xi^z+\mu^z_\oz\xi^\oz
\nonumber\\ c^\oz&=&(\exp i_\xi-1)(d\oz+\mu^\oz_zdz)=
 \xi^\oz+\mu^\oz_z\xi^z
\end{eqnarray}

This parametrization of the conformally invariant part of the metric and of the
ghosts is very useful to gauge-fix the conformally invariant part of the metric
since it  preserves automatically the factorization properties, and only
depends
on conformally invariant variables. Moreover, if one chooses
 a gauge where  the Beltrami differential is equal to a given background in the
holomorphic sector, the corresponding
ghost action  is simply
\be s(b_{zz}\m)= b_{zz}(\partial_\oz c^{z}+c^z\partial_z\mu^z_\oz-
\mu^z_\oz\partial_zc^z)
\ee In this gauge, the antighost $ b_{zz}$  is a quadratic differential with
BRST
transformation  $s b_{zz}=0$.  One has   similar expressions in the
anti-holomorphic direction, obtained by      changing   $\m$ and $\c$  into
$\mu^\oz_z$
and $c^\oz$. The conformal gauge is recovered for $\m=0$ and  $\m$ is the
source of the energy-momentum tensor components $T_{zz}$.  In the remaining of
this paper we will only consider the equations of the holomorphic sector.
The equations of the anti-holomorphic sector would follow simply  by
complex conjugation. These results which are specific
to 2-D reparametrization invariance have a natural explanation in the context
of
differential geometry \cite{bellon}. Note that the gauge fixing of global zero
modes can be done along these lines, as explained in \cite{global}

Let us see now that  the symmetry  equations can be written in a compact way,
by
unifying forms and ghosts. One defines
\be
\hat{\mu} ^z &=& dz + \mu^z_\oz d\oz + c^z  \nn\\
 \tilde d  &=&d+s
\ee  The curvature of $\hat{\mu} ^z$ is
\begin{equation}\hat {F}^{z}=(d+s){\hat\mu}^z - \frac {1}{2}\{{\hat\mu}^z,{\hat
\mu}^z\} =\tilde{d}\hat{\mu}^z-\hat{\mu}^z \partial_z\hat{\mu}_z
\end{equation} The BRST transformation laws defined in eq. (\ref{brs}) are just
the vanishing curvature condition
\begin{equation} \hat F^{z}=0
\end{equation}
One  should observe that $F^z=F_{\oz}dzd\z$ vanishes
identically, with no restriction on $\m$.

The existence and the expression of the   conformal anomaly under a factorized
form
  derive from the following  descent equations
 \begin{equation}
\hat{I}_4 = (s+d)\hat\Delta_3 = 0
\end{equation} with
\begin{equation}
\hat{\Delta}_3 = \hat{\Gamma}^z_z \tilde d \hat{\Gamma}^z_z =
\hat\mu^z\partial_z\hat\mu^z\partial^z_z\hat\mu^z
\end{equation} and
\begin{equation}
\hat\Gamma^z_z = \partial_z\hat\mu^z_z
\end{equation}
 One has indeed  $\tilde{d}
\hat{\Gamma}^z_z =\hat\mu^z  \p^2\hat \mu^z$ which implies that
$\tilde d  \tilde{\Gamma}^z_z \tilde d \hat{\Gamma}^z_z$
  vanishes identically  because $\hat\mu^z  \hat\mu^z =0$.
The consistent left conformal anomaly   is thus the
two-form component  with
 ghost number one of $\hat{\Delta}_3$. It can be written as \cite{bellon}
\be
  \Delta ^1_2=dzd\z(\p\c\p^2\m-\p\m\p^2\c)
\ee

      To obtain a conformal Lagrangian whose BRST
variation   reproduces the anomaly,   one defines  a
  Wess and Zumino scalar field $L$  with the following one form  field-strength
\begin{equation}
\hat G  = \tilde d L - \hat{\mu}\partial L - a \partial_z\hat{\mu}^z =\tilde d
L
- \{ \hat \mu, L \} - a \hat{\Gamma}^z_z
\end{equation}
  $a$ is any given  real number.
\def\d{\tilde d}
 This curvature satisfies the Bianchi identity
 \begin{equation}
\tilde d \hat G = \hat{\mu}^z\partial_z\hat G
\end{equation}
The  classical field-strength $G=G_{\z} d\z$ and  the  BRST  transformation
of the field L
are defined by
\be
\hat G =G_{\bar z}d\z
\ee
Indeed, the ghost decomposition of this equation is
\be
 G _{\bar z}=(  \partial _{\bar z} -\m \partial_z )L -a\partial_z \m
\ee
and
\be
\label{sL}  sL= \c\partial_z L - a\partial_z \c .
\ee
The last  equation   can be understood as the definition of the Wess and Zumino
field $L$.

The transformation law of the field strength $G=G_{\bar z}d\z$
 can be deduced directly from the Bianchi identity satisfied by $\hat G$
\begin{equation} sG_{\bar z} = c^z\partial _zG_{\bar z}
\end{equation}

The possibility of using $L$ as a Wess and Zumino field follows
from the equation
\def\p{\partial_z}
 \begin{eqnarray} a^2 \hat \Delta_3  = \tilde d (\hat \mu ^z  \partial _z  L
(\hat G +a\partial_z \hat \mu^z) -a  \hat G \partial _z \hat \mu ^z)
\end{eqnarray} This equation can be obtained by inserting the relation
  $a \hat{\Gamma}^z_z = \tilde d L - \hat{\mu}\partial L - \hat G  $ in the
expression of $\hat
\Delta_3$.
By expansion  in ghost number and form degree, one gets
 \begin{eqnarray}
 a^2\hat \Delta_3 &=& \tilde d \     [          \ (\p L(\partial_{\bar
z}-\m\p)L-2a\p\m)  \ dzd{\bar z} \nn\\& &+( \c\p L((\partial_{\bar
z}-\m\p)L-2a\p\m)  \nn\\& &\ -a\p\c ((\partial_{\bar z}-\m\p)L-a\p\m)+\m\p
L\big)\ d{\bar z}
 \nn\\ & &-  a\p L\p \c  \  dz -a  \partial _z  L \ \c\partial _z
\c \  ]      \end{eqnarray} The introduction of the field $L$ has
therefore the consequence of rendering trivial, i.e, equal to a sum of s-exact
and d-exact terms, all components  obtained from the ghost expansion of the
closed three-form
$\hat \Delta_3$.

One has in particular
 \begin{eqnarray} a^2 \int \Delta ^1_2&=& a^2 \int dz d\bar z \ \left [\partial
_z \c
\partial _z ^2\m -\partial _z \m \partial _z ^2\c\right]
\nn\\ &=&
\int dz d\bar z \ s\left[ \p L\left((\partial_{\bar
z}-\m\p\right)L-2a\p\m\right]
 \end{eqnarray} Therefore
\be
\label{wz} {\cal L}_{WZ}&=& \p L  \left((\partial_{\bar z}-\m\p)L-2a\p\m
\right) \nn\\
&=& \p L G_{\z}-a\p\m\p L\ee
 can be thought as  a Wess-Zumino Lagrangian density which can counterbalance
the
conformal anomaly.

Now, it is easy to verify that $\int dz d\bar z \ e^{-\frac{L}{a}}$ is an
invariant action, owing to the transformation law of $L$ defined in eq.
(\ref{sL}). Such a term is analogous to the cosmological term of the Liouville
action. It does not contribute to  the classical   energy momentum
tensor, since it is independent of $\m$.

 If, furthermore, we introduce a conformal field
$H_z$ with
\be  sH_z=\partial_z(c^zH_z)\ee
 we find that
\be
\int dz d\z \ H_z  G_{\z}
\ee
 is an invariant action.

Putting everything together, we are led to consider   the following action
 \begin{eqnarray}
\label{liou} \int dz  d\bar z \ [&&\left(H_z+\beta\p L\right)\left(( \partial
_{\z}-\mu^z_{\z}\partial_z)L- a \partial _z\mu ^z_{\bar z} \right)
 +\alpha
e^{-\frac{L}{a}}
\nn\\&& -\beta  a \partial _z\mu ^z_{\bar z}\partial _zL  \nn\\
&&- b_{zz}\left(\partial_\oz
c^{z }+c^z\partial_z\mu^z_\oz-
\mu^z_\oz\partial_zc^z \right)\ ]
 \end{eqnarray}

 The first term
ensures the propagation of $L$ and $H_z$ and the second one is a cosmological
type
 interaction.
The last term   is the ordinary $b-c$  ghost system resulting
from the gauge fixing of the Beltrami differential.  The terms which would
remain   for
$\beta=0$  are BRST invariant. The contributions of all terms proportional to
$\beta$     are such that their BRST variations reproduce
 the conformal anomaly. Both coefficients $a$
and
$\alpha$ can be set equal to one by field redefinitions and  one has the
freedom to chose at will  the value of the parameter
$\beta$    to compensate for a conformal anomaly,
resulting for instance from   a  coupling    to other fields.

 The energy-momentum tensor
is obtained by differentiating this action
 with respect to $\m$
\be T_{zz}&=& -(H_z+\beta\p L)\partial_zL+a\p H_z +2a\beta \p^2 L
\nn\\ & &+\p( b_{zz}\c )+b_{zz}\p\c
\ee Once again, it should be stressed that no reference to the conformal factor
of the metrics is necessary in this approach, and that mirror equations exist
in
the anti-holomorphic sector.

\section{  B-V approach to 2-D gravity in the Beltrami parametrization}
 We will
show now that the above action can be naturally interpreted in the B-V
formalism. We will closely follow the ideas introduced in \cite{previous} where
fields and anti-fields appear in dual combinations,   in a way which is
consistent with the unification of fields into forms, with a grading
equal to the sum of the ghost number (which is a
 negative integer for
the anti-fields) and of the ordinary form degree. In this section we will
consider
a minimal set of fields which   reproduces the results of the previous
section. In the next
 section we will see that new fields and their invariant action   can be
introduced if one
 pushes further our principle of unification.

Let us consider the     Beltrami differential and   ghost
  generalized one-form that we have already defined in the previous section
\begin{equation} \label{rest1}\hat \mu ^z =
 dz + \mu^z_{\bar z}d\bar z + c^z
\end{equation}
 According to \cite{previous}, it is natural
 to combine the anti-fields $M^{-1}_{zz}$ and $M^{-2}_{z z \bar z}$ of $\m$ and
$\c$ into    the following  generalized
    zero-form, "dual" to $\hat \mu ^z $
\begin{equation}
\label{rest2}
\hat M _z = M^{-1}_{zz}dz +M^{-2}_{z z \bar z}dz d\z
\end{equation}
One must also introduce the anti-fields $H^{-1}_{z\bar z}$ and $L^{-1}_{\bar
z}$ of the  Wess and Zumino sector
     fields
$L$ and $H_z$.     This leads us  to introduce the  following  generalized
zero-form and one-form, "dual" to each other
\begin{eqnarray}
\label{rest3}
\hat L &=& L^{-1}_{\bar z}d\z +L \nonumber \\
\hat H &=& H_zdz +H^{-1}_{z\bar z}dzd\bar z.
\end{eqnarray}

In the next section, we will show that other field components can occur in the
expansion of
the forms in eqs. (\ref{rest1}), (\ref{rest2})  and (\ref{rest3}).
In this sense, the pairs of fields and anti-fields
$(\m, M^{-1}_{zz}) $ and
$(\c, M^{-2}_{zz\bar z})$  in the pure gravity sector, and
$(L^{-1}_{\bar z}, H_z)$
and $(L, H^{-1}_{z\bar z})$ in the Wess and Zumino   sector, build up a minimal
system.

The curvatures of these generalized forms are
\begin{eqnarray}
\label{cons0}
{\cal F}^{z} &=& (s+d)\hat \mu ^z -\frac{1}{2}
  \{ \hat \mu ^z, \hat \mu ^z\}  = (s+d)\hat \mu ^z -
\hat \mu ^z \partial_z \hat \mu ^z
 \nonumber \\
 {\cal D}\hat M_z &=& (s+d)\hat M _z -
  \{ \hat \mu ^z, \hat M _z \}
 -\{ \hat H, L\}
\nn\\
&=&(s+d)\hat M _z -
\mu^z \partial_z \hat M _z +2\hat M _z\partial_z \mu^z -\hat H \partial_z \hat
L
\nn\\
 {\cal D}\hat L&=& (s+d)\hat L- \{ \hat \mu^z, \hat L \} = (s+d)\hat L -\hat
\mu^z \partial_z  \hat L
 \nn \\
 {\cal D}\hat H &=&  (s+d)\hat H-
\{ \hat \mu^z, \hat H \} =
 (s+d)\hat H -\partial_z(\hat \mu^z  \hat M )
\end{eqnarray}
These definitions are   consistent with Bianchi identities.

The  BRST symmetry    is then defined by the
following constraints on the curvatures
\begin{eqnarray}
\label{cons} {\cal F}^{z}  &=&0
\nonumber \\
 {\cal D}\hat M_z &=&a \partial_z\hat H
\nonumber\\ {\cal D}\hat L&=&a\partial_z \hat \mu ^z
 \nonumber \\
{\cal D}\hat H&=&0
\end{eqnarray}

  The explicit form of the action of $s$ on all fields and anti-fields is
obtained by expanding eqs. (\ref{cons}) in form degree and ghost
number. The property $s^2=0$ is the    consequence the Jacobi  relation
satisfied by
the graded bracket $\{ \ , \ \} $ appearing in eq. (\ref{cons0}), with
relations of
the type
\be {\cal F}^{z}={\cal D}{\cal D}=0\ee
 and
${\cal D}(\partial _z\tilde \mu ^z)=
\partial _z (\tilde d \hat \mu) - \{ \hat \mu , \partial _z
\hat \mu \} = 0$. One recovers of course the the same transformation laws of
the fields as in section (2).

This construction of the BRST symmetry is justified by its efficiency and also
by the fact that it follows the same pattern as for many other types of gauge
symmetries.

Let us denote generically all fields and ghosts by   $\phi$ and their
anti-fields by $\phi^*$. We will show that     the BRST    operation  defined
 in eq. (\ref{cons})  is associated to    the following  B-V action
\begin{eqnarray}
\label{bva0}
 S[\phi, \phi^*]= \int    \left[ \hat M_z( d \hat \mu ^z +\frac{1}{2}\{
\hat \mu , \hat \mu \}^z) +\hat H (d\hat L -\hat \mu \partial
\hat L
       -a\partial _z \hat \mu ^z)\right]^0_2
\end{eqnarray}Indeed, if one defines
\be {\  F}^{z(\hat \mu)} &=& d\hat \mu ^z -\frac{1}{2}
  \{ \hat \mu ^z, \hat \mu ^z\}
\ee and
\be
 G^{\hat L}=d\hat L -\hat \mu \partial \hat L
       -a\partial _z \hat \mu ^z
\ee one can rewrite the action (\ref{bva0}) as
\begin{eqnarray}
\label{bva}
 S[\phi, \phi^*]&=& \int    \left[\hat M_z {\  F}^{z(\hat \mu)}+\hat H
G^{\hat L}\right]^0_2
\end{eqnarray}
Then, by using the definition of $s$ given by eq. (\ref{cons}) one can verify
\be
\int   \ s\  \left(\hat M_z {\  F}^{z(\hat \mu)}+\hat H  G ^{\hat L}\right)=0
\ee
The component with ghost number one of this equation gives the wanted result
that
 the B-V action  (\ref{bva0}) is BRST invariant.

Reciprocally, the  action (\ref{bva0})  contains the information about the BRST
symmetry,
 through  the B-V equations
\be \label{defs} s\phi =  {\delta S[\phi, \phi^*] \over \delta
\phi^*}  \quad \quad s\phi^* =  {\delta S [\phi, \phi^*] \over \delta \phi}
\ee
 Let us verify this.   After expansion  in ghost number of all fields,  the
action  (\ref
{bva}) is
\begin{eqnarray}
\label{bv} S[\phi,\phi^* ]= \int dz d\z\    &[& H_z\left((\partial_{\bar z}
   -\mu^z_{\bar z}\partial _z)L-a\partial _z \mu ^z_{\bar z}\right)\nonumber \\
&
&+M^{-1}_{zz}(\partial _{\bar z}c^z +c^z\partial _z\mu ^z_{\bar z}-\mu
^z_z\partial _z c^z )\nonumber \\ & &+M^{-2}_{zz\bar z}c^z\partial _zc^z
   \nonumber \\ & &+H^{-1}_{z\z}(c^z\partial_zL-a\partial_z \c)\nonumber \\ &
&+L^{-1}_{\bar z} \partial _z(c^zH_z) \ \ ]
\end{eqnarray}
The
    action  (\ref{bv}) has a linear   dependance in the
anti-fields which means that it is of the first rank in the B-V sense. The BRST
transformations
of all fields with positive ghost number (the   quantum field theory
propagating
fields)    are thus the field polynomials which appear in     factor  of  the
anti-fields
 in eq.
(\ref{bv}).
It is then immediate to verify that if one applies  the relations  (\ref{defs})
to the
 action  (\ref{bv}), one gets    the same    expression  for  the action
 of  $s$ on all the fields and
anti-fields as the one obtained from  eqs. (\ref{cons}).

The B-V
procedure allows one to add a term proportional to  $ e^{-\frac{L}{a}} $ to
$S[\phi,
\phi^*]$ since such a term  is compatible with the symmetry.  This would modify
 the constraint
on the  curvature  of $\tilde L$ in eq. (\ref{cons})  by the addition of    the
equation of
motion stemming from this invariant term.    It is   also allowed to add  to
the B-V action  the Wess and Zumino
term  defined in eq. (\ref{wz})
 if one wishes to produce a theory  which
can compensate a conformal anomaly.

The   B-V formalism   indicates how one can  introduce    antighosts to perform
 the
gauge-fixing.  Since the  only ghost is $\c$, we have only  one possible gauge
function and thus only one possible antighost in the holomorphic sector. The
freedom in the
choice of the  gauge function   allows one  to introduce the antighost as a
quadratic
differential $b_{zz}$ in view of reaching for example the   gauge where $\m$ is
set equal to a background
value
$\mu^z_{\bar z 0}$. A  term
$b_{zz}^*\lambda^z_{\bar z}$ should be added to the action, where
$\lambda^z_{\bar z}$ is  a Nakanishi-Lautrup type Lagrange multiplier field and
$b_{zz}^*$ is the anti-field of $b_{zz}$. The gauge fixed action will be
obtained
by introducing the gauge function
 \be Z^{-1}=b_{zz}(\m-\mu^z_{\bar z 0})
\ee  and by replacing  all  anti-fields by mean of the constraint
 $\phi^*=\delta Z^{-1}/\delta  \phi$. As a result, the anti-field $M^{-1}_{zz}$
is set  equal to the usual antighost $b_{zz}$  and all
other anti-fields are zero. One eventually recovers the action defined in eq.
(\ref{liou}).

The s-transformation of the anti-field $M^{-1}_{zz}$ is
\begin{eqnarray} sM^{-1}_{zz} =\frac{\delta {\cal L}_{z\z}}{\delta \mu _{\bar
z}^z}=T_{zz}
     \end{eqnarray} From our point of view, this equation explains the fact
that
 the energy-momentum tensor $T_{zz}$ is a $Q$-commutator
in
the Hamiltonian formalism \cite{verlinde}.

Apart from technical details, the  interesting result of this section is the
simplicity of the B-V action defined in eq. (\ref{bva}).
 This   action, which contains the whole information about
the transformation laws of the field of 2-D gravity including the Wess and
Zumino sector, is analogous to  a Chern-Simon   action. This  is not too
much  a surprise in view of earlier  results,  where the relevance of the
equation
$\hat\mu\d \hat\mu=0$  had been
emphasized   for building the BRST algebra of conformal models \cite{bellon}.
This relationship  is probably related to the
topological nature of the ghost and  Wess and Zumino sector sector of string
theory.

Let us conclude this section by indicating   how our formulae
permit a straightforward derivation  of
$2-D$ topological gravity equations  \cite{gravite}. Following ref.
\cite{previous}, one introduces the topological ghosts of 2-D gravity  as the
components with positive ghost number of a 2-form
$\hat X^z_2=\Psi^z_{\bar z}d\z+\Phi^z$. The associated  anti-fields are then
the
components of a dual   $(-1)$-form $\hat Y_{-1 z}=
\Psi ^{*-2}_{ z z}d z+\Phi^{*-3}_{ \z z z}dzd\z $. The invariant B-V
action is then
 \begin{equation} \label{BVT} \int    \left[  \hat M_z(  F^{z\hat \mu}+\hat
X^z_2)+
 \hat X^z_2 D^{\hat\mu} \hat Y_{-1 z})\right]^0_2
\end{equation}
   The   transformation laws for the fields and anti-fields which leave
invariant this action are defined by the curvature constraints
 \begin{eqnarray}
\hat{\cal F}^z
  &=&\tilde d \hat \mu
     -{1\over 2} \{ \hat \mu , \hat \mu\} = \hat X ^z_2
     \nonumber \\
     {\cal D}\hat M_z
  &=&\tilde d \hat M_z
   - \{ \hat \mu , \hat M _z \}
     = \{ \hat X ^z_2, \hat Y_{-1 z}\}
     \nonumber \\
     {\cal D}\hat  X^z_2
  &=&\tilde d \hat X^z_2
   - \{ \hat \mu , \hat X ^z_2 \} = 0 \nonumber \\
   {\cal D}\hat Y_{-1 z}&=&\tilde d
\hat Y_{-1 z}
   - \{ \hat \mu , \hat Y_{-1 z} \} = 0
\end{eqnarray} These formulae give
 the BRST transformation laws of the topological 2-D gravity   in the
Beltrami parametrization as in ref.
\cite{singergravity}. With suitable choices of gauge functions, the gauge
fixing of the B-V action (\cite{{BVT}}) would reproduce  the
known actions for 2-D topological gravity.

\section{A more general action for 2-D gravity with Wess and Zumino terms }

We have just found how to incorporate in a rather simple
algebraic framework all fields and anti-fields relevant to 2-D gravity,
including the Wess and Zumino sector. Our basic tools have been    the Beltrami
parametrization of the conformally invariant part of the metric and   the
unification of fields and anti-fields into  forms graded by   the sum of their
ordinary form degree and   ghost number.  The latter quantity is positive   for
the
ordinary ghosts and negative for their anti-fields, which explains why some
components of the
forms have higher ordinary form degree.

However,    by looking at  eqs. (\ref {rest1}, \ref {rest2},
\ref {rest3}),  we see  that we have restricted our-selves in the expansion of
forms since
$\hat\mu^z$, $\hat M_z$, $\hat L$ and
 $\hat H_z$, which are respectively generalized 1-form,   0-form,  0-form and
1-form,   could
contain additional  components matching the grading requirements. When we have
defined the BRST
symmetry by imposing constraints on the curvatures of these forms, these
restrictions have
yield  no contradiction because of the identity
$d\mu^z-\mu^z\p\mu^z=0$.  It is thus quite natural to generalize the field
contents by
considering instead of eqs. (\ref {rest1}, \ref {rest2},
\ref {rest3})  the following  general  form decomposition
 \def\m1{\mu^{-1}_{\z}}
\be\label{total}
 \label{final}\tilde \mu ^z &=& \m1 dzd\z+\nu dz   +(dz + \mu^z_{\bar z}d\bar
z)
+ c^z
\nn\\
 \tilde M _z &=&M_z+ M^{-1}_{z\z}d\z+ M^{-1}_{zz}dz +M^{-2}_{z z \bar
z}dzd\z\
\nn\\
\tilde L &=& L+L^{-1}_{\z} d\z+ L^{-1}_z dz+L^{-2}_{z\z} dz d\z
\nn\\
\tilde H &=& H^1+H_{\z} d\z+H_z dz +H^{-1}_{z\z}dzd\z
\end{eqnarray}

Let us   look at the components with   positive ghost number in these
expansions.
There are three  new   classical fields with ghost number zero, namely
$\nu$  which has conformal weight zero,
 and $M_z$ and $H_z$  which have   conformal weights one. Then, there is $H^1$
which has
ghost number one, and which represents   an additional gauge freedom.

Thus, besides the   pairs of fields  and anti-fields of the previous section
$(\mu^z_{\z}, M^{-1}_{zz})$,
$(\c, M^{-2}_{zz\bar z})$,    $( H_z, L^{-1}_{\bar z} ) $ and
$(L, H^{-1}_{z\bar z})$, we have now the pairs
$(\nu, M^{-1}_{z\z})$, $(M_z, \mu^{-1}_{\z})$, $(H_{\z}, L^{-1}_{z})$  and
$(H^1, L^{-2}_{z\z})$.
 The introduction of the object $\nu dz$ in the
expansion of the Beltrami differential      will  imply a      non
vanishing  value for classical component of the Beltrami curvature.

\def\m{{\mu^{z}_{\z}}}

The action of the BRST symmetry on all the fields is given by the
same curvature constraints   as in the previous section. The property
$(s+d)^2=0$ still holds,  since the new fields have been introduced  just as
new
components of the   differential forms   (\ref{total}) and the constraints are
compatible with the Bianchi identities. One has therefore
\begin{eqnarray}
\label{constotal} {\cal F}^{z} &=& (s+d)\tilde \mu ^z -\frac{1}{2}
  \{ \tilde \mu ^z, \tilde \mu ^z\}
\nn\\ &=& (s+d)\tilde \mu ^z -
\tilde \mu ^z\partial_z\tilde \mu ^z=0 \nonumber \\ {\cal D}\tilde M_z &=&
(s+d)\tilde M _z -
  \{ \tilde \mu ^z, \tilde M _z \}
 -\{ \tilde H,\tilde L\}
\nn\\ &=&(s+d)\tilde M _z -
\mu^z \partial_z \tilde M _z +2\tilde M _z\partial_z \tilde \mu ^z -\tilde H
\partial_z \tilde L =a \partial_z\tilde H
\nonumber
\\ {\cal D}\tilde L&=& (s+d)\tilde L- \{ \tilde \mu^z, \tilde L \}
\nn\\ &=& (s+d)\tilde L -\tilde \mu^z \partial_z  \tilde L
   =a\partial_z  \tilde \mu ^z \nonumber \\ {\cal D}\tilde H&=&  (s+d)\tilde H-
\{ \tilde \mu^z, \tilde H \}
\nn\\ &=&
 (s+d)\tilde H -\partial_z(\tilde \mu^z  \tilde H )=0
\end{eqnarray}
The   corresponding       B-V
action is the same as in eq. (\ref{bva}),  except that the forms have now a
more
general decomposition in the fields and anti-fields.     Using the
decomposition given by  eq. (\ref{total}), one gets the following B-V action
\begin{eqnarray}
\label{bvaa} S[\phi,\phi^*]= \int dzd\z\  &[&
M_z\left((\pa_{\z}-\m\p)\nu+\nu\p\m\right)
\nn\\
 & &
+H_z((\pa_{\z}-\m\p)L-a\p\m)
\nn\\
 & & +H_{\z}(a\pa_{z}\nu+\nu\p L)
\nn\\
 & &  +M^{-1}_{zz}(\partial _{\bar z}c^z +c^z\partial _z\mu ^z_{\bar z}-\mu
^z_{\z}\partial _z c^z )
\nonumber \\ & & +M^{-1}_{z\z}
(\c\p\nu-\nu\p\c)
\nn\\ & &
+\mu^{-1}_{\z}(\c\p M_z+2M_z\p\c-H^1\p L-a\p  H^1)
\nn\\
& &+M^{-2}_{zz\bar z}c^z\partial _z c^z
\nn\\
 & &+H^{-1}_{z\z}(c^z\p  L-a\p \c)
\nonumber \\
 & &+L^{-1}_{\z} (\p (c^z H_z-\nu H^1)
\nonumber \\
 & & +L^{-1}_{z}\left((\pa_{\z}-\m\p)H^1-H^1\p\m+\p(\c H_{\z})\right)
\nn\\
 & & +L^{-2}_{z\z}\p(\c H^1)\ ]
\end{eqnarray}
 Let us consider the   classical part of this action
\begin{eqnarray}
\label{bvaaa} S[\phi,\phi^*=0]= \int dzd\z\  &[&
 M_z((\pa_{\z}-\m\p)\nu+\nu\p\m)\nn\\  && + H_z((\pa_{\z}-\m\p)L-a\p\m) \nn\\
&&
+H_{\z}(a\pa_{z}\nu+\nu\p L ) \ \ ]
\end{eqnarray} It is invariant under the following   symmetry
\be s\m &=& \pa_{\z}\c+\c\p\m-\m\p\c
\nn\\ s\nu&=&\c\p\nu-\nu\p\c
\nn\\ sM_z&=&\c\p M_z+2M_z\p\c-H^1\p L-a\p  H^1
\nn\\ sH_z&=&\p(\c H_z-\nu H^1)
\nn\\ sH_{\z}&=&(\pa_{\z}-\m\p)H^1-H^1\p\m+\p(\c H_{\z})
\nn\\ sL&=&\c\p L-a\p\c
\ee
We see that in addition to the reparametrization invariance governed by the
ghost $\c$ we have another gauge symmetry governed by the ghost $H^1$, with $s
H^1=\p(\c H^1)$ and that
$H_{\z}$ plays the role of a gauge field for  this symmetry.  It is
quite interesting that the action can be written as
\begin{eqnarray} S[\phi,\phi^*=0]= \int dzd\z\left[
 M_z F_{\z}
 +H_zG_{\z}  +H_{\z}G_{z}\right ]
\end{eqnarray} with
\begin{eqnarray} F_{\z}&=&(\pa_{\z}-\m\p)\nu+\nu\p\m \nn\\
 G_{\z}&=&(\pa_{\z}-\m\p)L-a\p\m  \nn\\
 G_{z}&=& \nu \pa_{\z}L+a\p L
\end{eqnarray} The invariance of the action can be verified from
\begin{eqnarray} sF_{\z}&=&\c\p F_{\z}-F_{\z}\p\c\nn\\ s G_{\z}&=&\c\p G_{\z}
\nn\\ s G_{z}&=&\c\p G_{z}
\end{eqnarray} and there are non trivial compensations between the variations
of
the three terms of the action which involve the ghost $H^1$.

For a further clarification of this formula, let us notice that we can
summarize  in the following compact way the transformation laws of $\m$ and
$\nu$
\be
\label{last}\tilde F^z&=&(s+d)(\nu dz + dz+\m  d\z+\c)\nn\\ & &-{1\over2}\{ \nu
dz
+ dz+\m  d\z+\c,\nu dz + dz+\m  d\z+\c\} \nn\\ & &=F^z=F_{\z}dzd\z
\ee and
\be\label{last1} (s+d)\tilde F^z=\{ \nu dz + dz+\m  d\z+\c,\tilde F^z\}
\ee The relation  $sF_{\z}=\c\p F_{\z}-F_{\z}\p\c$ is the component with ghost
number one of eq. (\ref{last1}). There are similar formulae which involve
$G_{ z}$ and $G_{\z}$.

A very direct way to prove the invariance of the action is to define
\be {  F}^{z\tilde \mu} &=& d\tilde \mu ^z -\frac{1}{2}
  \{ \tilde \mu ^z, \tilde \mu ^z\}
\ee and
\be
  G^{\tilde L}=d\tilde L -\tilde \mu \partial \tilde L
       -a\partial _z \tilde \mu ^z
\ee and to check
\be
\int  \   s  \left[ \tilde M_z {\  F}^{z(\tilde \mu)}+\tilde H   G^{\tilde L}
\right] =0
\ee The proof is quite straightforward from the relations
\be d{   F}^{z\tilde \mu}=\{\tilde \mu^z, {   F}^{z\tilde \mu }\}
\ee
\be d   G^{\tilde L} =\{ \tilde \mu ^z,   G ^{\tilde L}\} +\{L , {
F}^{z\tilde \mu} \}
-a\p{  F}^{z\tilde \mu}
\ee

The introduction of the field $\nu$
   implies  therefore   that one has a non vanishing  classical     Beltrami
curvature  $ F^{\z}=\left((\pa_{\z}-\m\p)\nu+\nu\p\m\right)dzd\z$, contrarily
to what happens in the usual case which involves a restricted number of fields.

 Another
distinction  is the existence of a new degree of gauge freedom associated to
the
ghost   $H^1$. This freedom can be used for instance to gauge-fix to zero the
field
$H_{\z}$.  To do so one  introduces an antighost  ${\bar r}_z$ associated to
$H^1$, its anti-field
$r_{\z}^*$ and the associated  Lagrange multiplier field  $\beta_z$, with
$s{\bar r}_z=\beta_z$. Then one  adds to the B-V
action
 the  term $r
_{\z}^*
\beta_z$.
The gauge fixed action is finally obtained
by considering the gauge function
 \be Z^{-1}=b_{zz}(\m-\mu^z_{\bar z 0}) +{\bar r}_z H_{\z}
\ee  By  replacing all  anti-fields in the B-V action (\ref{bvaa}) by
$\phi^*=\delta Z^{-1}/\delta  \phi$, one obtains the following   action
\be
\label{bvaaaa}  \int dzd\z\  &[&
 M_z\left((\pa_{\z}-\m\p)\nu+\nu\p\m\right)
\nn\\
 & &
 +H_z\left((\pa_{\z}-\m\p)L-a\p\m\right)
\nn\\
 & &
 -b_{zz}({\pa_{\z}\c +\c\p\m-\m\p\c})
\nn\\
 & &
 -{\bar
r}_z\left((\pa_{\z}-\m\p)H^1-H^1\p\m+\p(\c H_z) \right)\ \ ]
\end{eqnarray}
 This action is of the conformal type.
In addition to the
usual propagating pairs $b_{zz}-\c$  and $  H_z-L$ we have now the pairs $
M_z-\nu$ and
${\bar r}_z-H^1$.
 The last term in the action
 is the  ghost term  corresponding to the gauge-fixing to zero of $ H_{\z}$.
Other types of gauge-fixing for the field
$H_{\z}$ could be defined  which would lead us to different ghost interactions
for $H^1$.

The form of the consistent anomaly $\tilde\Delta _3$, with
$(d+s)\tilde\Delta _3=0$ is unchanged, since the basic structure equations are
the same, and we have not found an anomaly in the   symmetry parametrized by
the
ghost $H^1$. The same Wess and
Zumino Lagrangian density as in the restricted theory is thus applicable to
this model to make
it anomaly free.

Our construction has therefore led us to propose
the following action for the  2-D gravity, which includes a Wess and Zumino
term
and a cosmological type term
\begin{eqnarray}
\label {lioufin} S_{2D}= \int dz d\bar z \  &[&
(H_z+\beta\p L)(
(\partial_{\z}-\m\partial_z)L- a \p\m )
\nn\\
& &-\beta  a \partial _z\mu ^z_{\bar z}\partial _zL+\alpha
e^{-\frac{L}{a}}
\nn\\ & &+
 M_z((\pa_{\z}-\m\p)\nu+\nu\p\m)
 \nn\\
& & -b_{zz}(\partial_\oz
c^{z}+c^z\partial_z\mu^z_\oz-
\mu^z_\oz\partial_zc^z)
\nn\\
 & & -{\bar r}_z\left((\pa_{\z}-\m\p)H^1-H^1\p\m+\p(\c H_z) \right) \ ]
 \end{eqnarray} The associated energy momentum tensor is
\be T_{zz}&=& -(H_z+\beta\p L)\partial_zL+a\p H_z +2a\beta \p^2 L\nn\\ & &
-M_z\p\nu-\p(M_z\nu) \nn\\ & & +\p( b_{zz}\c )+b_{zz}\p\c\nn\\ & &
+\p({\bar r}_z H^1)
\ee
These expressions for the Lagrangian and energy momentum tensor    should be
complemented by their mirror  expressions, obtained by   complex conjugation.
$a$ and $\alpha$ can be set equal to one as in the   case
with  the restricted  set of
fields, while     $\beta$ can be choosen at will, possibly with different
values in both
holomorphic and anti-holomorphic   sectors.

\section{conclusion}
 We have  applied the B-V formalism for 2-D
gravity with a  Wess and Zumino  sector. By using    the Beltrami
parametrization
of conformal field theories,  we have
found      the same type of unification between all fields and anti-fields
 as the one  we had previously observed in \cite{previous} for   the theories
of forms coupled to Yang-Mills fields. Moreover, we have shown that the B-V
action has a structure quite similar  to that of   a Chern-Simon   action. We
have  introduced
       new conformal fields    with a conceptually very simple
action.  In addition to  the ordinary conformal invariance, this  action has a
new gauge
symmetry and induces new ghost interactions, with a possible assymmetry between
the
holomorphic and anti-holomorphic sectors.  Its  properties   and its  possible
couplings
to matter will be studied in a separate publication.

 \vskip 2cm
\noindent {\bf { Acknowledgments: }} {The author would like to
express his deep gratitude to RIMS for the hospitality
extended to him during his stay in Japan.}


%


\end{document}